\documentclass[12pt]{article}
\usepackage{graphicx}
\begin{document} 

\title{Potts-like model for ghetto formation in multi-cultural 
societies}

\author{Christian Schulze\\
Institute for Theoretical Physics, Cologne University\\D-50923 K\"oln, Euroland}

\maketitle
\centerline{e-mail: ab127@uni-koeln.de}

\bigskip
Abstract:
In a Potts-like model of $Q$ ethnic groups, we follow Schelling
(1971) and Meyer-Ortmanns (2003) and simulate the formation
of ethnic ghettos as well as their prevention by an increasing
social temperature.  

\bigskip


\bigskip

\section{Introduction}
            
Binary models like Ising-type simulations have a long history.
They have been applied by Schelling \cite{schelling} to describe
the ghetto formation in the inner cities of the USA, i.e. to
study phase separation between black and white. More recently, 
Meyer-Ortmanns simulated the Ising model with a temperature
increasing with time and showed that in this way ghettos can
be avoided: Higher temperature means higher tolerance towards 
different people \cite{ortmanns,weidlich}. The present note aims to 
generalize this work to up to seven different ethnic groups, as 
may be more appropriate to European societies. 

\begin{figure}[hbt]
\begin{center}
\includegraphics[angle=-90,scale=0.45]{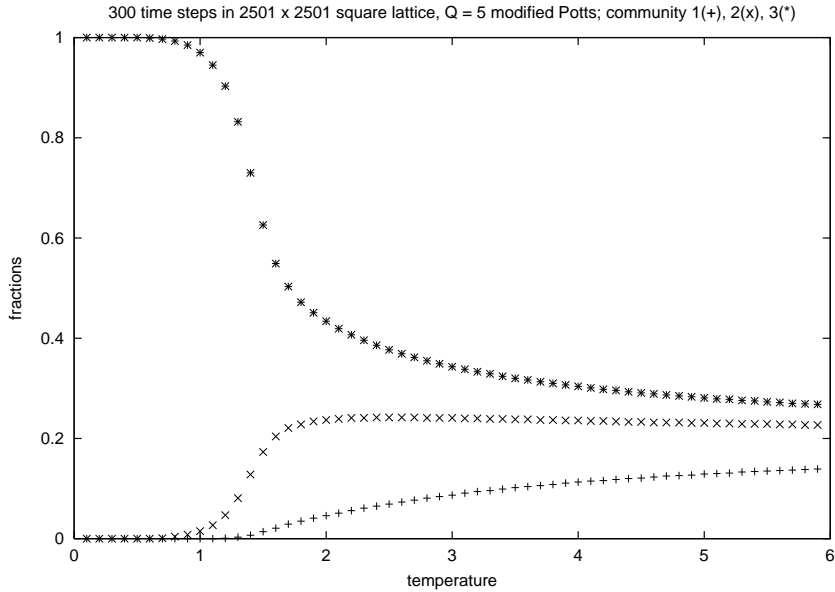}
\end{center}
\caption{Equilibrium concentrations of five ethnic groups, 
starting with group 3 and allowing transformation of members
of one group into members of another group.}
\end{figure}

\begin{figure}[hbt]
\begin{center}
\includegraphics[angle=-90,scale=0.45]{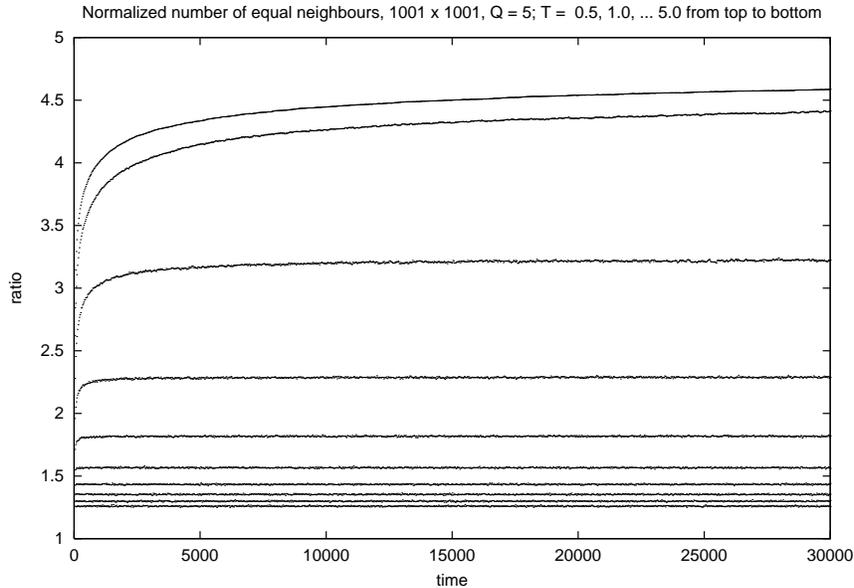}
\end{center}
\caption{Time dependence of separation for five groups in
one million lattice sites; the initial value (random distribution)
is normalized to unity. }
\end{figure}

\begin{figure}[hbt]
\begin{center}
\includegraphics[angle=-90,scale=0.3]{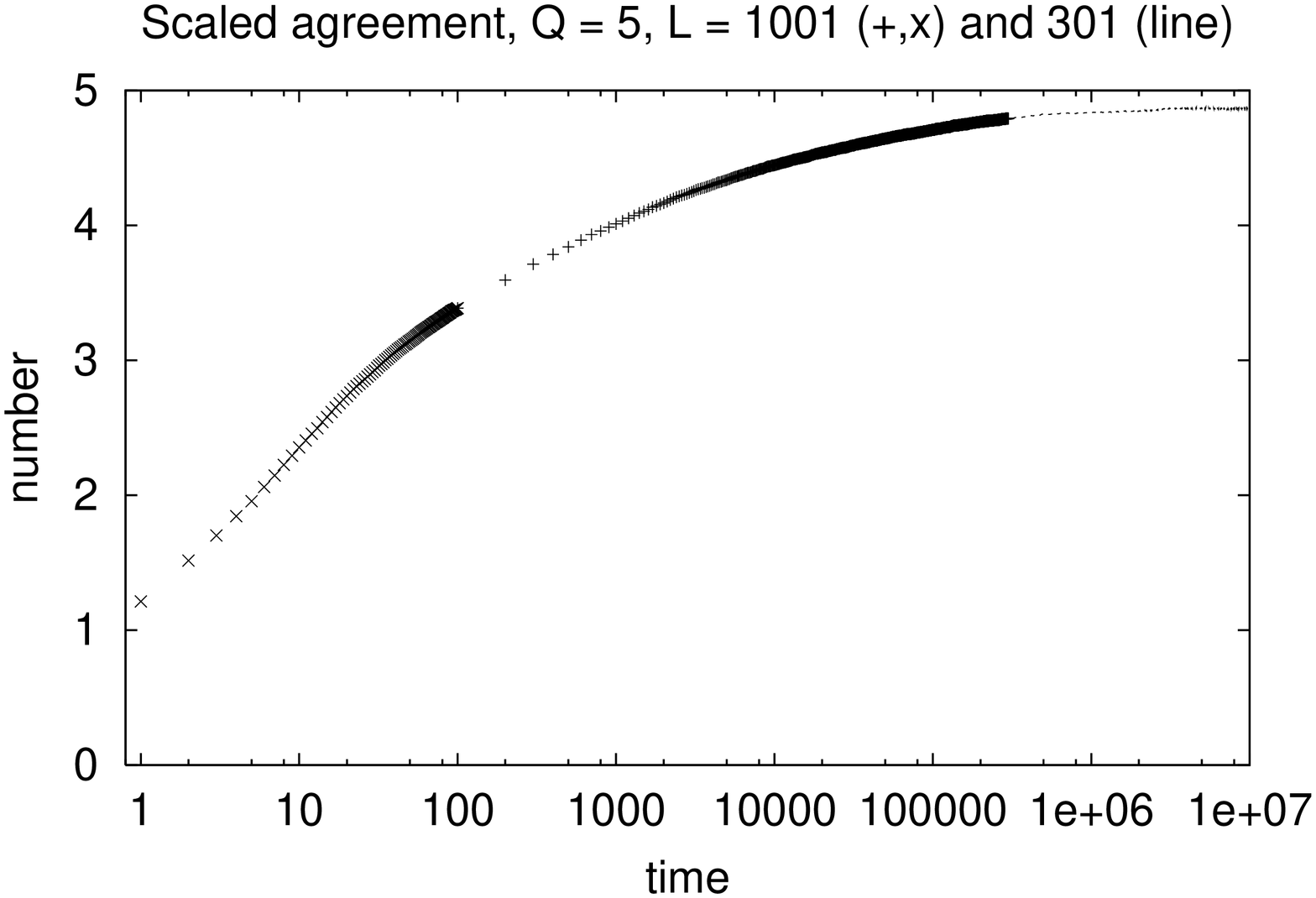}
\includegraphics[angle=-90,scale=0.3]{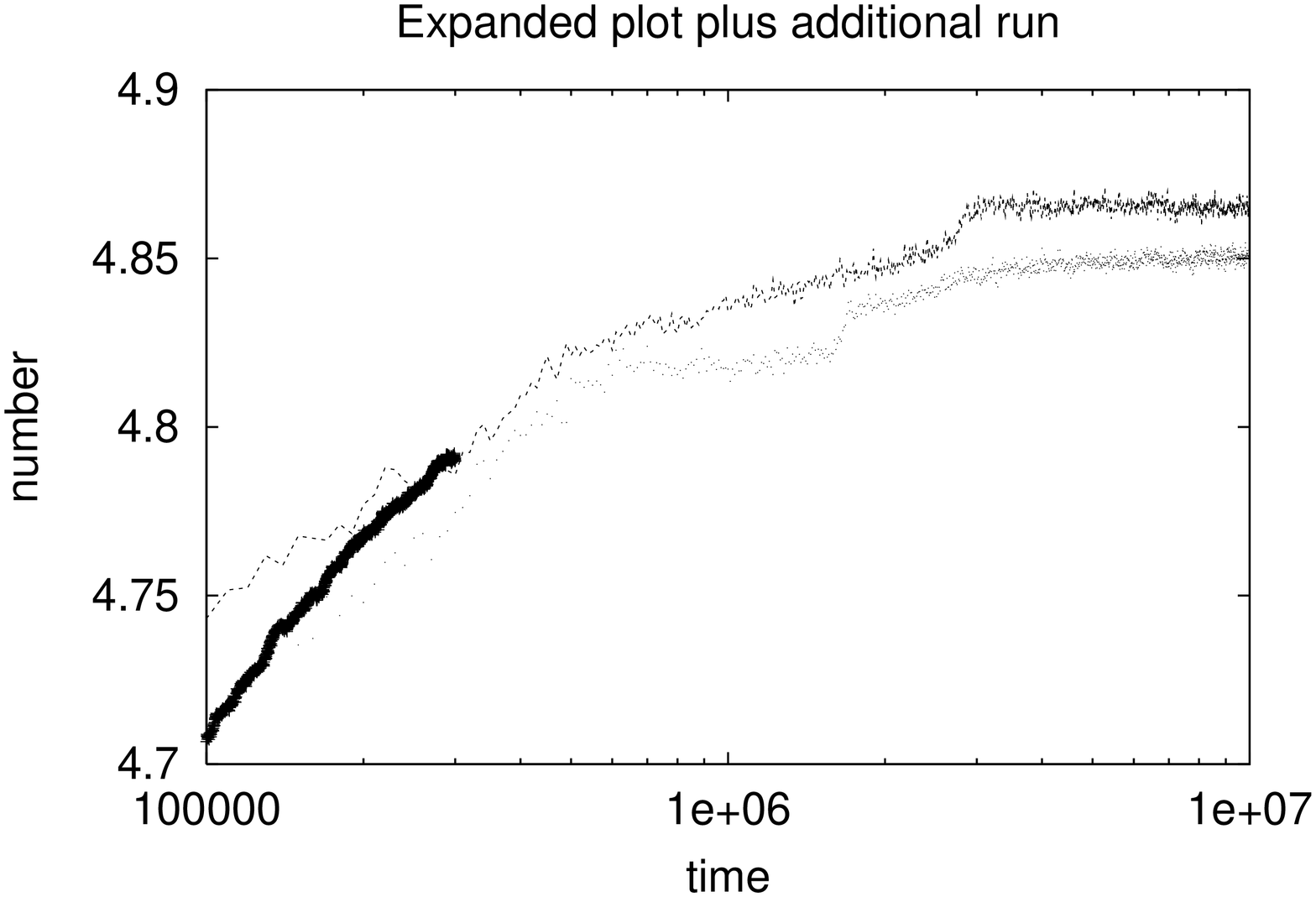}
\end{center}
\caption{a: As Fig.2 but for longer times and $T = 0.5$ only. b:
Same run as in part a, plus another run differing only
by the random numbers, shown only for the later times. }
\end{figure}

\begin{figure}[hbt]
\begin{center}
\includegraphics[angle=-90,scale=0.5]{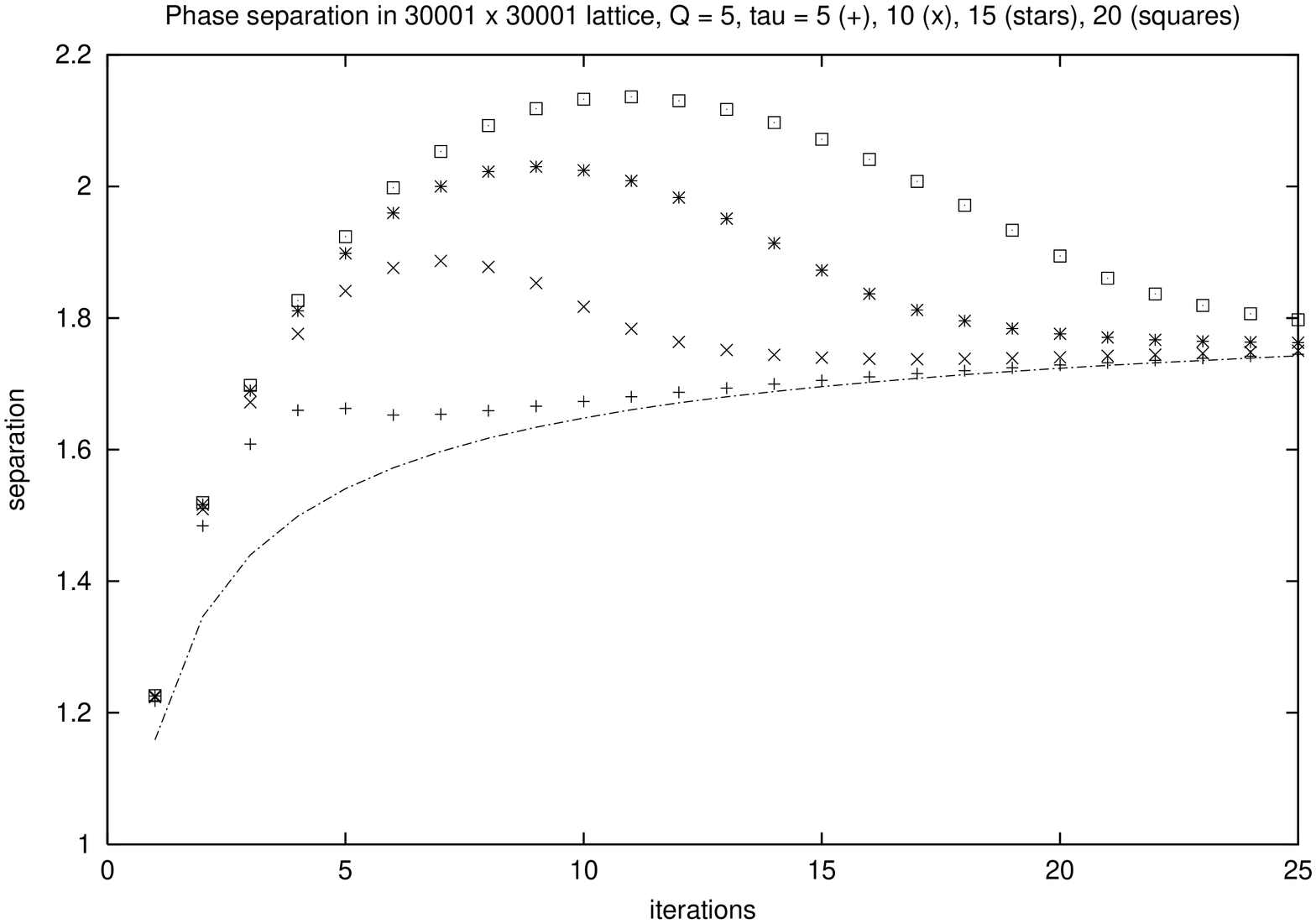}
\end{center}
\caption{Time dependence of separation for five groups
in $30001 \times 30001$ lattice with temperature increasing
from 0.5 to 2.5 in the first $\tau$ iterations, with $\tau=5,
10,15,20$  from bottom to top. The line shows the behaviour 
for fixed temperature $T = 2.5$ as in Fig.2. For $1001 \times 1001$
the deviations are of the order of the symbol size. The long-time
value from Fig.2 is 1.82.
}
\end{figure}

\section{Model}

We assume the presence of $Q$ different ethnic groups, numbered
by an index $q$ between 1 and $Q$. The similarities between 
$q$ and $q \pm 1$ are taken as larger than those between groups
with indices which are far apart. Thus, in contrast to the usual 
Potts  model, we take the interaction energy between two groups
$i$ and $k$ as proportional to the absolute value of the 
difference $q_i-q_k$: 
$$E = J \sum |q_i -q_k| \eqno (1)$$ 
where the sum goes over all nearest neighbours on a $L \times
L$ square lattice, with helical bounday conditions.
Different configurations are realized with a 
Boltzmann probability $\propto \exp(-E/k_BT)$ and the Boltzmann
constant $k_B$; the temperature $T$ thus is controlled by the
dimensionless variable $k_BT/J$. However, we measure our 
social temperature in units of the critical Potts temperature 
[2/ln$(1+\sqrt q)]J/k_B$. Thus, in the Ising case of $Q=2$ 
the equilibrium critical temperature is $T=1$ since with two
groups the difference between our interaction energy and the 
usual Potts energy vanishes. 

For equilibrium studies the details of the dynamics do not
matter much and we use Glauber dynamics; for non-equilibrium
instead we use Kawasaki dynamics with infinite-range exchange.
Thus in the latter case each person randomly selects a site far 
from its own neighbourhood and tries to exchange its residence
with the person on that site, according to the above Boltzmann
probability. 

\section{Results}

While the $Q=2$ case (not shown) has a sharp phase transition at 
$T=1$, with mixing for $T > 1$ and phase separation for $T< 1$,
for higher $Q$ (we simulated up seven cultures) no sharp 
phase transition exists. Fig.1 shows a typical case, $Q = 5$:
At low temperatures, $q = 3$ dominates, at high temperatures,
all $q$ are roughly equally represented, without a sharp 
separation temperature in between. (Group 4 has about the same fraction as
group 2, and group 5 agrees in size with group1.) 

For the dynamics, we no longer allow members to change the 
group to which they belong. To quantify the separation
effects, we calculate the nearest-neighbour correlation 
function, i.e. the average number of equal neighbours which
a site has. We normalize this number by its value in the 
initial configuration, where all groups are equally and randomly 
distributed among the lattice sites. This normalized number
of equal neighbours is shown in Fig.2 for $Q=5$ and various 
temperatures $T$: The higher is $T$, the lower is the separation
as measured by the number of equal neighbours. Low temperatures
show an unusual long-time dynamics, as shown in Fig.3 for 
$T = 0.5$.

Following Meyer-Ortmanns \cite{ortmanns} we now assume that
the tolerance towards other ethnic groups, as measured through
the temperature $T$, increases with time. Thus we start with 
$T=0.5$ which means according to Figs.2,3 a strong separation.
Then, starting from the first iteration, we increase $T$ linearly
with time up to $T = 2.5$ (weak separation according to Fig.2)
within $\tau$ iterations. We check if the 
resulting non-equilibrium separation (measured in the same units
as Fig.2) remains below a threshold of two, i.e. is at most 
twice as high as for randomly distributed people. Fig.4 with  
900 million people shows that for $\tau = 5$ the separation
increases monotonically up to its equilibrium value near 1.8,
for $\tau =10$ it has a maximum below 1.9, for $\tau =15$  the
maximum is near 2.0, and for $\tau =20$ it is clearly above 2.
Thus tolerance has to increase fast enough if separation is
to be avoided.

Thanks are due to D. Stauffer and M. Hohnisch for help and discussions.

\end{document}